\documentclass[journal=jacsat,manuscript=article]{achemso}
\usepackage[version=3]{mhchem} 
\usepackage{siunitx}
\usepackage[dvipsnames]{xcolor}
\usepackage{amssymb}
\usepackage{pifont}
\usepackage{wrapfig}
\usepackage{hyperref}

\usepackage{amsmath} % or simply amstext

\usepackage{tabularx,colortbl}
\usepackage{capt-of}
\usepackage{adjustbox}
\usepackage[font=normalsize,labelfont={bf},font={bf}]{caption}

\author{Chakradhar Guntuboina}
\affiliation[ece]
{Department of Electrical and Computer Engineering, Carnegie Mellon University, 15213, USA}
\alsoaffiliation[contribution]
{Authors Contributed Equally}

\author{Adrita Das}
\affiliation[biomed]
{Department of Biomedical Engineering, Carnegie Mellon University, 15213, USA}
\alsoaffiliation[contribution]
{Authors Contributed Equally}

\author{Parisa Mollaei}
\affiliation[meche]
{Department of Mechanical Engineering, Carnegie Mellon University, 15213, USA}

\author{Seongwon Kim}
\affiliation[cheme]
{Department of Chemical Engineering, Carnegie Mellon University, 15213, USA}

\author{Amir Barati Farimani}
\email{barati@cmu.edu}
\affiliation[meche]
{Department of Mechanical Engineering, Carnegie Mellon University, 15213, USA}
\alsoaffiliation[biomed]
{Department of Biomedical Engineering, Carnegie Mellon University, 15213, USA}
\alsoaffiliation[mld]
{Machine Learning Department, Carnegie Mellon University, 15213, USA}

\title[An \textsf{achemso} demo]
{PeptideBERT: A  Language Model based on Transformers for Peptide Property Prediction}

\begin{document}

\begin{abstract}

\noindent Recent advances in Language Models have enabled the protein modeling community with a powerful tool since protein sequences can be represented as text. Specifically, by taking advantage of Transformers, sequence-to-property prediction will be amenable without the need for explicit structural data. In this work, inspired by recent progress in Large Language Models (LLMs), we introduce PeptideBERT, a protein language model for predicting three key properties of peptides (hemolysis, solubility, and non-fouling). The PeptideBert utilizes the ProtBERT pretrained transformer model with 12 attention heads and 12 hidden layers. We then finetuned the pretrained model for the three downstream tasks. Our model has achieved state of the art (SOTA) for predicting Hemolysis, which is a task for determining  peptide's potential to induce red blood cell lysis. Our PeptideBert non-fouling model also achieved remarkable accuracy in predicting peptide's capacity to resist non-specific interactions. This model, trained predominantly on shorter sequences, benefits from the dataset where negative examples are largely associated with insoluble peptides. Codes, models, and data used in this study are freely available at: \href{https://github.com/ChakradharG/PeptideBERT}{https://github.com/ChakradharG/PeptideBERT}

\end{abstract}

%\clearpage
\section{Introduction}

Peptides are organic molecules containing amino acids, ranging from only a few amino acids to numerous units that are joined together in ordered sequences\cite{langel2009introduction, damodaran2008amino, degrado1988design, voet2016fundamentals, bodanszky2012principles, Mollaei2023.04.16.536913}. The length and arrangement of amino acids in a sequence govern a protein's structural and biological properties\cite{schulz2013principles, petsko2004protein, mollaei2023activity, yadav2022prediction}. Consequently, peptide sequence determines how the peptide engages with its environment and various molecules. For example, the peptides' therapeutic properties such as hemolysis, fouling characteristics, and solubility\cite{varanko2020recent, dunn2015peptide, schueler2017modeling} are defined by sequences of amino acids. Hemolysis refers to the disintegration of red blood cells\cite{ponder1948hemolysis}, and understanding its connection to the peptide's amino acid sequence is vital for formulating safe and efficacious peptide-based treatments. Peptides that are fouling are less likely to adhere to or interact with molecules present in their environment\cite{harding2014combating, yu2011anti}. By exploring the influence of the peptide sequence on non-fouling properties, one can engineer bio-compatibility, durability, and overall effectiveness of designed biomaterials, medical devices, and drug delivery systems. Peptides' solubility which refers to the ability of a peptide to dissolve in a solvent significantly affects their delivery and efficacy\cite{sarma2018peptide}. Understanding and manipulating this sequence-structure-function relationship is crucial for peptide design in drug development and biomolecular engineering\cite{fosgerau2015peptide}. Given the significance of mapping the sequence of peptide to its properties, there have been many modeling attempts to perform this task. The Quantitative Structure-Activity relationship (QSAR) models were previously used to build the relationship between sequence and structural properties of chemical compounds\cite{QSAR2014modelling}. QSAR were used to predict the properties of several classes of peptides to sequences including inhibitorypeptides\cite{newquantitativestructure,ACE-I,ACE-1study}, antimicrobial peptides\cite{Gramnegative,Gramnegative2,Gramnegative3} and anti-oxidant peptides\cite{antioxidants,antioxidants1,antioxidants2}. For solubility predictions, DSResol\cite{DSResol} outperformed models such as DeepSol\cite{DeepSol}, SoluProt\cite{Soluprot}, Protein-Sol\cite{proteinsol} with an accuracy of 75.1\%. However, most of these models require structure of the peptide, which is difficult to have access for a large variety of peptides. DSResol discerns extensive-range interaction information among amino acid k-mers utilizing dilated convolutional neural networks. MahLooL\cite{MahLooL} has comparable performance with respect to DSResol. MahLooL outperforms DSResol only for peptides of very short length (18-50), with an accuracy of 91.3\%. MahLooL employs bidirectional Long Short-Term Memory (LSTM) networks to capture extensive sequence correlations. HAPPENN\cite{HAPPENN} stands as a state-of-the-art (SOTA) model for predicting hemolytic activity, achieving an accuracy of 85.7\%. HAPPENN employs normal features selected through Support Vector Machines (SVM) and an ensemble of Random Forests. 

With the rise of Transformers and Large Language Models (LLMs)\cite{vaswani2017attention, devlin2018bert, brown2020language}, new deep learning architectures have emerged for modeling protein sequences since amino acid sequences can be considered as words and sentences similar to the language. Specifically, the attention mechanism of LLMs allows them to capture both immediate and intricate connections between elements of various types of textual data. As a result, it has initiated a revitalization in the field of bioinformatics since protein sequences, similar to languages, exhibit complex interactions among amino acids. Using LLM and Transformers, we are now able to leverage advanced language modeling techniques to investigate the contributions of amino acids in the protein's features. In this study, and by taking advantage of Transformers and pretraining, we developed PeptideBERT, a language model that predicts the peptide properties using only amino acid sequences as the input.  By taking advantage of pretrained models such as ProtBert, we fine-tuned PeptideBert to be able to predict the peptide's properties.(Figure \ref{fig:model}) Pretrained models such as ProtBert \cite{protbert} learned the protein sequence representation by being trained on massive protein sequences. We demonstrated that PeptideBERT can predict the hemolysis, non-fouling characteristics, and solubility of a given peptide using language models. 

\section{ Methods}

\subsection{Datasets}
\begin{figure}[t!]
     \centering
     \includegraphics[width=1.0\linewidth]{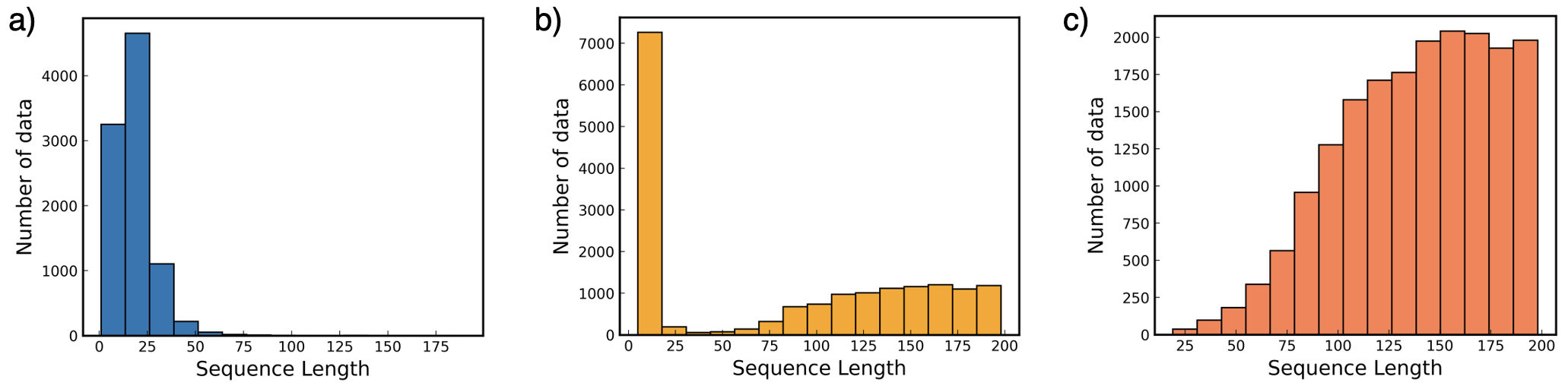}
     \caption{Sequence Length of each Peptide property dataset (a) Hemolysis, (b) Non-fouling and (c) Solubility.} 
     \label{fig:barplot}
 \end{figure}

The datasets employed for each specific task and their corresponding sequence length distributions are visually depicted in  Figure \ref{fig:barplot}. For the non-fouling dataset, the length of sequences falls within the range of 2 to 20 residues. This particular dataset focuses on comparatively shorter sequences, likely to capture specific characteristics relevant to the non-fouling property. In contrast, the dataset utilized for the solubility task encompasses a broader spectrum of sequence lengths, spanning from 18 to 198 residues. This wide-ranging sequence length distribution is indicative of the diverse nature of sequences included in this dataset, potentially accommodating a variety of structural and functional attributes. The use of datasets with distinct sequence length profiles highlights the tailored approach taken to address the unique requirements of each predictive task, further enhancing the model's ability to capture and interpret the relevant information accurately.

\begin{figure}[t!]
     \centering
     \includegraphics[width=1.0\linewidth]{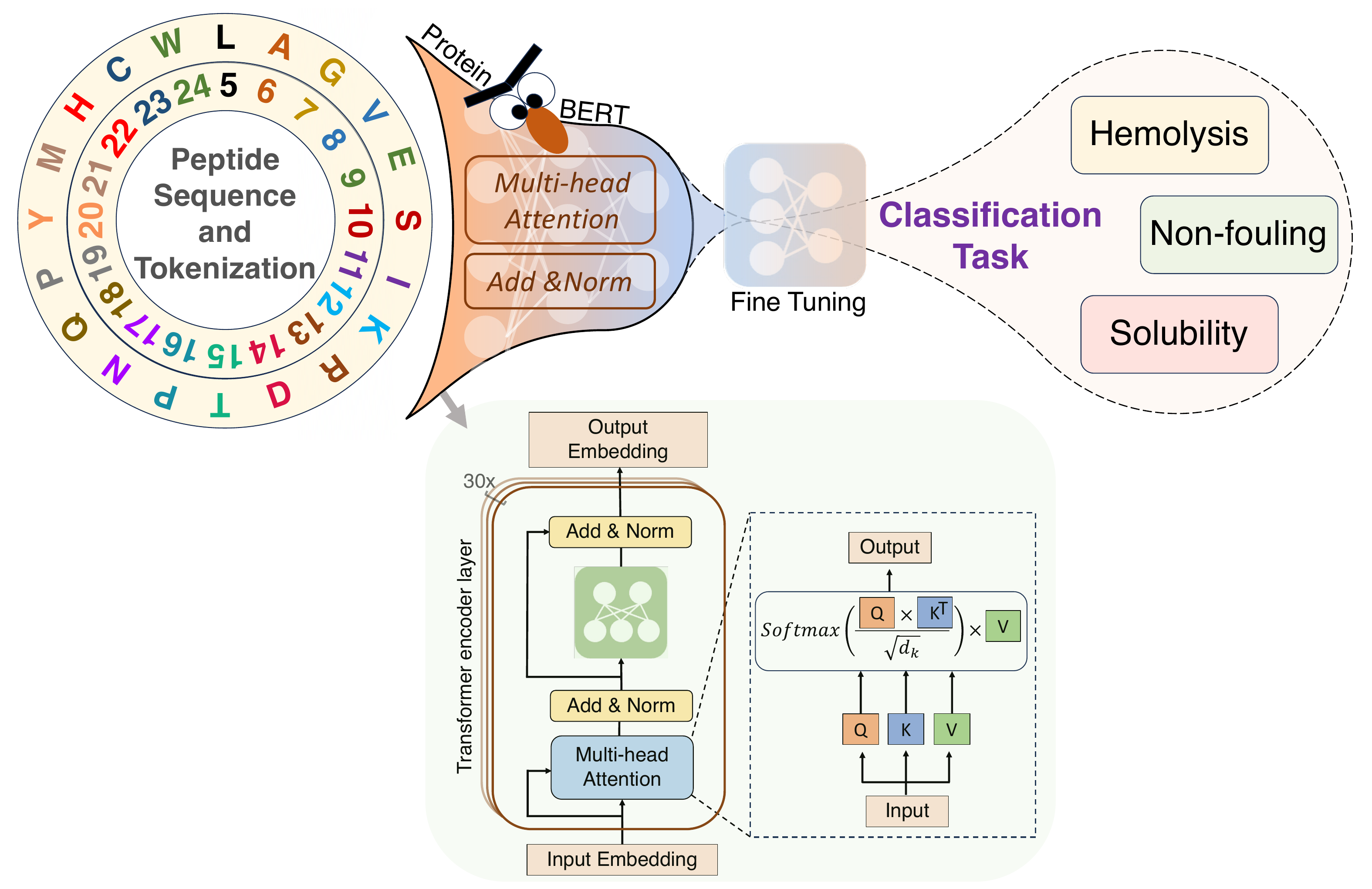}
     \caption{ The model architecture of PeptideBERT. Peptide sequences are tokenized and subsequently processed through ProtBERT. Subsequently, a classification head of Multi-Layer Perceptrons (MLPs) is added for fine-tuning process. The model is individually trained on three different classification downstream tasks: Hemolysis, Non-fouling, and Solubility} 
     \label{fig:model}
 \end{figure}

\subsubsection{Hemolysis}
The term \textbf{hemolysis} relates to the disruption of the membranes of red blood cells, which leads to a decrease in the lifespan of cells. It is essential to identify antimicrobial agents or peptides that do not cause hemolysis, as this ensures their safe and non-toxic use against bacterial infections. Antimicrobial peptides (AMPs) represent a collection of small peptides, recognized for their efficacy against bacteria, viruses, fungi, and even cancer cells. Although these peptides exhibit limited bio-availability and short lifetime, they possess distinct advantages over other categories of drugs or peptides including their notable specificity, selectivity, and minimal toxicity\cite{AMPs, AMPs1}. However, distinguishing between peptides that cause hemolysis and those that do not is challenging because their main effects occur on the charged surface of bacterial cell membranes. Primarily, these peptides function as agents that modulate the immune response, induce apoptosis, and hinder cell proliferation. In recent times, there have been various endeavors to compile databases containing AMPs and to use computational techniques to categorize their hemolytic properties. In this study,  the Database of Antimicrobial Activity and Structure of Peptides (DBAASPv3)\cite{DBAASP} was utilized for the hemolytic activity prediction model. The extent of activity is assessed by extrapolating measurements from dose-response curves to the point where 50\% of red blood cells (RBCs) undergo lysed. Peptides with activity below 100 $\mu$g/mL are categorized as hemolytic\cite{MahLooL}. Each measurement is treated as an independent case meaning that sequences can appear multiple times in the dataset. The training dataset comprises 9316 sequences, with 19.6\% being positive (hemolytic) and 80.4\% being negative (non-hemolytic). The sequences consist only of L- and canonical amino acids. It is worth noting that due to the inherent variability in experimental data, around 40\% of observations contain identical sequences that are labeled as both negative and positive. For instance, a sequence like "RVKRVWPLVIRTVIAGYNLYRAIKKK" has been found to exhibit both hemolytic and nonhemolytic behavior in two different laboratory experiments, resulting in two distinct training examples\cite{MahLooL}.

\subsubsection{Solubility}
The solubility dataset consists of 18,453 sequences, with 47.6\% being labeled as positives and 52.4\% as negatives. These labels are based on information sourced from PROSO II\cite{PROSO}. The solubility of the sequences was determined through a retrospective evaluation of electronic laboratory notebooks, which were part of a larger initiative known as the Protein Structure Initiative. The analysis involves tracking the sequences through various stages (such as Selected, Expressed, Cloned, Soluble, Purified, Crystallized), HSQC (heteronuclear single quantum coherence), Structure determination, and submission in the Protein Data Bank (PDB)\cite{PRSISGKB}. The categorization of peptides as soluble or insoluble is explained in PROSO II\cite{PROSO},achieved by contrasting their experimental status at two specific time points, September 2009 and May 2010. Specifically, those proteins that were initially insoluble in September 2009 and remained in the same insoluble state eight months later were classified as insoluble.

\subsubsection{Non-fouling}
The information employed to forecast resistance against nonspecific interactions (non-fouling) is gathered from reference 40\cite{nonfouling}. The positive dataset comprises 3,600 sequences, while the negative examples are drawn from 13,585 sequences, yielding a distribution of 20.9\% positives and 79.1\% negatives. The negative data are drawn from insoluble and hemolytic peptides, along with scrambled positives. To generate the scrambled negatives, sequences are chosen with lengths drawn from the identical range as their corresponding positive set. The residues for these sequences are chosen based on the frequency distribution observed in the solubility dataset. To address the class imbalance stemming from the disparity in dataset size for negative examples, which can lead to the model being biased towards the majority class and performing poorly on the minority class, the samples are assigned weights to indicate the importance of each example during training. The dataset was compiled following the approach outlined in ref 41\cite{C2SC21135A}. A non-fouling peptide (considered a positive example) is defined following the methodology introduced by White et al\cite{C2SC21135A}. White et al. demonstrated that the amino acid frequencies on the exterior surfaces of proteins differ significantly, with this discrepancy becoming more pronounced in environments prone to protein aggregation, such as the cytoplasm. They established that synthesizing self-assembling peptides adhering to this amino acid distribution and applying these peptides to surfaces yields non-fouling surfaces. This pattern was also observed within chaperone proteins, an area where mitigating nonspecific interactions is crucial\cite{WHITE20122484}.

\subsection{Data Preprocessing}
The provided datasets\cite{MahLooL} have been preprocessed by applying a custom encoding method. In this encoding, the integer representation of each of the 20 amino acids is given by its index in the following array (indexing starts from 1) : [A, R, N, D, C, Q, E, G, H, I, L, K, M, F, P, S, T, W, Y, V]. For example, the sequence `A M N D V' is converted into 1 13 3 4 20.

Since we are using ProtBERT and since its tokenizer uses a different encoding process, to ensure compatibility, we first converted all the datasets from integers into characters using reverse mapping and then converted them back into integers using ProtBERT's encoding. Following the encoding procedure, we split each dataset into 3 non-overlapping subsets, a training set (to train the model) consisting of 81\% of the dataset, a validation set (for hyperparameter tuning) consisting of 9\% of the dataset, and a test set (to benchmark the model's performance on unseen data) consisting of 10\% of the dataset. This specific train-validation-test split of 81\%-9\%-10\% has been selected to ensure a proper comparison between our approach and the previous methodologies\cite{MahLooL}. Data augmentations, if any (such as in the \textit{Solubility } task), are then applied to the training set while the validation and test sets remain unchanged.

\subsection{Data Augmentation}
The following data augmentation techniques were applied to the solubility dataset in order to improve the model's classification accuracy on the task:
\begin{itemize}
    \item random\_replace: Randomly replace a given fraction of the unpadded protein sequence with random amino acids. For example, if the fraction is 0.1, and the protein sequence is `A M N D V E T R L H', then the output will be something like `A M N D V E \textbf{M} R L H'.
    \item random\_delete: Randomly delete a given fraction of the unpadded protein sequence. For example, if the fraction is 0.1, and the protein sequence is `A M N D V E T R L H', then the output will be something like `A N D V E T R L H'.
    \item random\_replace\_with\_A: Randomly replace a given fraction of the unpadded protein sequence with the amino acid `A'. For example, if the fraction is 0.1, and the protein sequence is `A M N D V E T R L H', then the output will be something like `A M N \textbf{A} V E T R L H'.
    \item random\_swap: Randomly swap a given fraction of adjacent pairs of amino acids in the unpadded protein sequence. For example, if the fraction is 0.1, and the protein sequence is `A M N D V E T R L H', then the output will be something like `A M N D V E T R \textbf{L R} H'.
    \item random\_insertion\_with\_A: Randomly insert amino acid `A' into the unpadded protein sequence and subsequently increase its length by a given fraction. For example, if the fraction is 0.1, and the protein sequence is `A M N D V E T R L H', then the output will be something like `A M N D V E T R L \textbf{A} H'.
    \item random\_mask: Randomly mask or replace certain elements in the sequence with a [MASK] token. The mask token is typically chosen to represent missing or irrelevant information and is often assigned a specific integer value. If the masking probability is 0.2, about 20\% of the elements in the sequence will be selected for masking. For example, if given a protein sequence `A M N D V E T R L H', after masking the selected elements with [MASK] token, the sequence becomes `A M \textbf{[MASK]} D \textbf{[MASK]} E \textbf{[MASK]} R L H'.
\end{itemize}

Applying these augmentations resulted in varying degrees of improvement in the model's classification accuracy. The results are shown in Table \ref{tab:augmentation}. The best performing augmentation was random\_swap with a 0.843\% increment in accuracy.

\subsection{Model Architecture}
The architectural blueprint of PeptideBERT is given in Figure \ref{fig:model}. At its core, PeptideBERT uses the pretrained ProtBERT\cite{protbert}, a transformer model that consists of 12 attention-heads and 12 hidden layers. Its design is influenced by the original BERT\cite{bert} model. ProtBERT is pretrained on a massive corpus of protein sequences (UniRef100\cite{uniref100}) containing over 217 million unique protein sequences. During its pretraining phase, a Masked Language Modelling (MLM) objective was employed. Here, 15\% of the amino acids in sequences were masked, challenging the model to predict these hidden segments based on the surrounding context. Additionally, this pretraining was performed in a self-supervised manner, using only raw protein sequences without any human-generated labels. The attention mechanism is a pivotal component of transformer architectures, designed to model dependencies in sequences irrespective of the distance between elements. At its core, the attention mechanism computes a weighted sum of input values (often termed `values' or V), where each weight indicates the relevance or `attention' a specific input should receive given a query. The weights are determined by calculating the dot product between the query (Q) and associated keys (K), followed by a softmax operation to ensure the weights are normalized and sum to one. This allows the transformer to focus more on certain parts of the input while attending less to others. In the context of Natural Language Processing, for instance, this can mean focusing on specific words in a sentence that are more pertinent to understanding the context or meaning of another word. The multi-head attention architecture further enhances this by enabling the model to attend to multiple parts of the input simultaneously, capturing diverse relationships in the data. By doing so, transformers can learn intricate patterns and long-range dependencies, making them particularly effective for a plethora of sequence-based tasks. Such a transformative encoder structure in ProtBERT allows the model to glean context-sensitive representations of amino acids, treating each protein sequence akin to a 'document'. ProtBERT is followed by a regression head, which is a fully connected neural network that takes the output of ProtBERT and maps it to a continuous value. The regression head is a single fully connected layer with 480 nodes. The output of the regression head is passed through a Sigmoid function to ensure that the output is between 0 and 1. The output of the Sigmoid function is then thresholded at 0.5 to obtain the final binary prediction. The optimal architecture for the regression head was determined by performing a series of experiments, the results of which are discussed in the Results section.

\subsection{Training Procedure}
For each task, a separate model was fine-tuned on the corresponding dataset. The model was trained using the \textit{AdamW} optimizer of binary cross-entropy loss function with an \textit{initial learning-rate} of 0.00001 and a \textit{batch size} of 32. The model was trained for 30 epochs. \textit{ReduceLROnPlateau } scheduler was employed to reduce the learning rate by a factor of 0.1 if the validation accuracy did not improve for 4 epochs. The model was trained on a single NVIDIA GeForce GTX 1080Ti GPU with 16GB of memory. The training time and the optimal hyperparameters for each task are outlined briefly in the \textbf{supporting information }.

\section{Results and Discussion}
The performance and efficiency of our proposed model, PeptideBERT is shown through a comprehensive analysis of its achieved outcomes. The \textit{Solubility } prediction task presented a significant challenge due to the presence diverse range of length of sequences within the dataset. Given the complexity and variability of peptide sequences, this particular prediction task demanded a tailored approach to enhance the model's performance. To address this challenge and improve the model's ability to generalize across a wide spectrum of sequences, we employed an augmentation strategy. Table \ref{tab:augmentation} outlines the various augmentation techniques we applied and their impact on \textit{Solubility} prediction accuracy. This approach aimed to expose the model to a more comprehensive array of sequence variations, effectively expanding its learning capacity. By performing augmentation on the dataset, we were able to introduce increased diversity of sequence patterns and characteristics, enabling the model to better capture the underlying features that influence solubility prediction. \textit{Random replace} at a rate of 2\% led to an accuracy of 68.694\%, while random delete, also at 2\%, yielded an accuracy of 68.814\%. The introduction of \textit{Random replace with A} at 2\% demonstrated an accuracy of 68.573\%. Notably, \textit{Random swap} augmentation at 2\% showcased an improved accuracy of 70.018\%.

\begin{table}[h!]
    \begin{tabular}{ |c|c|c| } 
    \hline
    Augmentations applied & Train set size & Accuracy(\%) \\
    \hline
	random\_replace(2\%) & 29892 & 68.694 \\
	random\_delete(2\%) & 29892 & 68.814 \\
	random\_replace\_with\_A(2\%) & 29892 & 68.573 \\
	random\_swap(2\%) & 29892 & 70.018 \\
	random\_insertion\_with\_A(2\%) & 29892 & 69.597 \\
	random\_swap(2\%), random\_insertion\_with\_A(2\%) & 44838 & 68.453 \\
	random\_swap(2\%), random\_insertion\_with\_A(1\%) & 44838 & 68.814 \\
	random\_swap(3\%) & 29892 & 68.814 \\
	random\_replace\_with\_A(2\%), random\_insertion\_with\_A(2\%) & 44838 & 69.054 \\
    \hline
    \end{tabular}
    \caption{\label{tab:augmentation}Ablation results for different augmentation techniques for \textit{Solubility} prediction. Baseline accuracy (without any augmentations) is 69.175\%}
\end{table}

Similarly, \textit{Random insertion with A} at 2\% exhibited an accuracy of 69.597\%. A combination of \textit{Random swap} and \textit{Random insertion with A}, both at 2\%, achieved an accuracy of 68.453\% on a larger training set of 44838 samples. It is interesting to note that employing a lower rate (1\%) of \textit{Random insertion with A} in conjunction with \textit{Random swap} maintained an accuracy of 68.814\%. The application of \textit{Random swap} at 3\% resulted in an accuracy of 68.814\%, akin to the accuracy produced by \textit{replacing with A} and \textit{inserting with A}, both at 2\%. Table \ref{tab:acc} provides a comprehensive comparison of classification accuracies accross various models, including our novel ProtBERT based model across the three distinct prediction tasks. For the \textit{Non-fouling} prediction task, our PeptideBERT model demonstrated exceptional performance, achieving an accuracy of 88.365\%, significantly surpassing the accuracy of 82.0\% attained by the Embedding + LSTM approach.

\begin{table}[h!]
    \begin{tabular}{ |l|c|c| } 
    \hline
    Approach & Task & Accuracy(\%) \\
    \hline
    PeptideBERT (Ours) & Non-fouling & \textbf{88.365} \\
    Embedding + LSTM & Non-fouling & 82.0 \\
    \hline
    PeptideBERT (Ours) & Hemolysis & \textbf{86.051} \\
    Embedding + Bi-LSTM & Hemolysis & 84.0 \\
    UniRep + Logistic Regression & Hemolysis & 82.0 \\
    UniRep + Random Forests & Hemolysis & 84.0 \\
    HAPPENN\cite{HAPPENN} & Hemolysis & 85.7 \\
    HLPpred-Fuse\cite{HLPpred} & Hemolysis & - \\
    one-hots + RNN\cite{onehots} & Hemolysis & 76.0 \\
    \hline
    PeptideBERT (Ours) (With Augmentation) & Solubility & 70.018 \\
    PeptideBERT (Ours) (Without Augmentation) & Solubility & 69.175 \\
    Embedding + Bi-LSTM & Solubility & 70.0 \\
    PROSO II\cite{PROSO} & Solubility & 71.0 \\
    DSResSol (1)\cite{DSResol} & Solubility & \textbf{75.1} \\
    \hline
    \end{tabular}
    \caption{\label{tab:acc}Classification Accuracy Comparison of previous methods and our ProtBERT based approach on each of the 3 prediction tasks}
\end{table}

Moreover, the PeptideBERT model outperformed the other models in the \textit{Hemolysis} task, achieving an accuracy of 86.051\%, while the Embedding + Bi-LSTM and UniRep + Logistic Regression approaches achieved 84.0\% and 82.0\% accuracies, respectively. This showcases the robustness of our model in predicting hemolytic properties. In the \textit{Solubility } prediction task, our PeptideBERT model demonstrated competitive results. With data augmentation, it achieved a predictive accuracy of 70.018\%, while without augmentation, it attained an accuracy of 69.17\%. Comparatively, the Embedding + Bi-LSTM and PROSO II methods achieved 70.0\% and 71.0\% accuracies, respectively. These findings highlight the effectiveness of our PeptideBERT-based approach, which consistently achieved higher accuracies across all three prediction tasks, both with and without data augmentation, showcasing its potential to enhance predictive capabilities in diverse bioinformatics applications.

\begin{figure}[h!]
     \centering
     \includegraphics[width=1.0\linewidth]{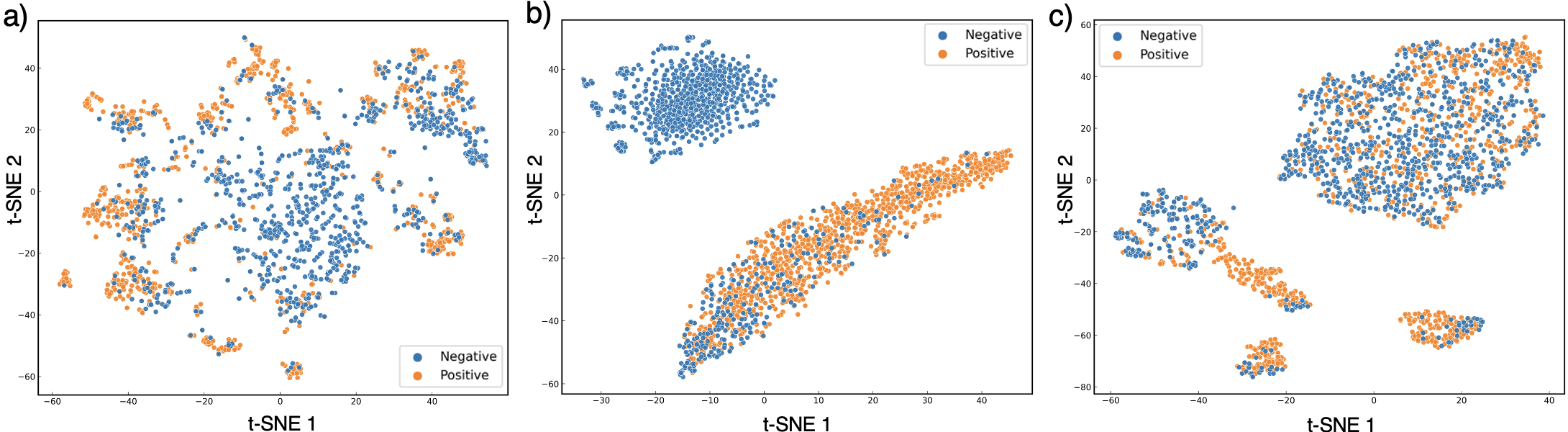}
     \caption{t-SNE visualization of peptide properties (a) Hemolysis, (b) Non-fouling and (c) Solubility. The [CLS] token embedding from the last hidden state of PeptideBERT is visualized after dimensionality reduction.}
     \label{fig:tsne}
 \end{figure}
 
 Transformer's attention mechanism enables every token embedding in the encoder to capture the information of the whole input sequence. However, in practical applications, the classification token ([CLS] token) often serves as a comprehensive representation of the entire sequence \cite{schwaller1}\cite{schwaller2}. For effective classification, the ProtBERT tokenizer adds [CLS] token at the start of each sequence, thus, enabling this token to contain the information from all token embeddings. Using this insight, to effectively visualize PeptideBERT's understanding of various peptide sequences and its classification competence, we extracted the [CLS] tokens of each peptide sequence from the final hidden state and visualized with t-distributed Stochastic Neighbor Embedding (t-SNE)\cite{Maaten1990tsne}. The t-SNE algorithm evaluates pairwise similarities in the high-dimensional space attracting similar data points toward each other while repelling dissimilar points. The visualization results of [CLS] token embeddings which have a size of (480) are shown in Figure \ref{fig:tsne}. The t-SNE visualization clearly illustrates that peptides with similar properties (represented by identical color markers) are clustered together. Furthermore, the result indicates the model's capability of classifying peptides solely based on their sequence information and that the [CLS] token within the embeddings has effectively captured the distinguishing features of individual peptides. From the observed patterns, the model appears to segregate peptides into two distinct groups (outer and inside for the Hemolysis dataset) meaning that the binary classification downstream tasks were valid for fine-tuning the PeptideBERT. The model's errors are also noticeable, for example, the blue dots positioned in the bottom right of the Non-fouling t-SNE \ref{fig:tsne}(b) means the misclassified peptides as negative Non-fouling. Comparing all three plots, the Hemolysis \ref{fig:tsne}(a) and Non-fouling \ref{fig:tsne}(b) t-SNE shows clear classification while the Solubility t-SNE has a relatively large number of errors aligning with the accuracy results from the fine-tuning procedure \ref{tab:acc}.

In this paper, we introduced three sequence-based classifiers aimed at predicting peptide hemolysis, solubility, and resistance to nonspecific interactions(non-fouling). These classifiers demonstrate competitive performance in comparison to the latest state-of-the-art models. The PeptideBERT model for hemolysis prediction task is designed to predict a peptide's capacity to cause red blood cell lysis. It is tailored for peptides spanning 1 to 190 residues and involves L- and canonical amino acids. PeptideBERT provides state-of-the-art sequence-based hemolysis predictions with an accuracy of 86.051\%. This accuracy suggests that the model can reliably identify peptides with hemolytic potential, contributing to better decision-making in peptide design and application. It is important to note that the training dataset (refer to \textbf{Datasets} section for a brief outline) for hemolysis prediction comprises peptide sequences that possess antimicrobial or clinical significance. While this targeted approach certainly boosts the model's performance within these particular domains, prudent consideration is needed when extending its predictions to a wider array of peptides. The fact that our hemolytic model is specifically designed for peptides with lengths ranging from 1 to 190 residues reflects an important consideration that, peptide length can significantly influence their behavior, including interactions with cells or molecules. By tailoring the model to this specific length range, it takes into account the structural variations that can arise in different peptide lengths. Our non-fouling model also provides state-of-the-art predictions with an accuracy of 88.365\% for the non-fouling task, which is designed to predict the ability of a peptide to resist non-specific interactions. The training data for the non-fouling task primarily consists of shorter sequences in the range of 2-20 residues. The dataset employed for this task consists of instances of negative examples that are predominantly associated with insoluble peptides, which could lead to an increase in accuracy if only soluble peptides are compared\cite{MahLooL}. Our predictive model achieves an accuracy of 70.018\%, with augmentations, for the solubility task. This accuracy can be primarily attributed to the challenges involved in predicting solubility in cheminformatics\cite{MahLooL}.

\section{Conclusion}
In this work, we developed a language model called PeptideBert to predict various peptide properties including hemolysis, solubility, and non-fouling. Our model takes advantage of pretrained models that learnt the representation of protein sequences. Using PeptideBert, we demonstrate a hemolysis predictor and a non-fouling predictor that outperforms existing state-of-the-art models. The performance of these classifiers demonstrates their potential utility in the field of peptide research and applications. Notably, the model for hemolysis prediction exhibits robust predictive capabilities, offering valuable insights into the potential of peptides to cause red blood cell lysis. However, it is important to acknowledge the focused nature of its training dataset, which primarily encompasses sequences with antimicrobial or clinical relevance. As such, while these classifiers show promising results, a prudent approach involves considering the context and potential limitations when applying their predictions to a broader range of peptides. The competitive results compared to state-of-the-art models underline the progress made in predictive peptide modeling using language models. It suggests that the newly introduced models are not just novel, but also effective in capturing relevant features that contribute to peptide behavior. The predictive capabilities of these classifiers hold promise for diverse applications, ranging from drug design to bioengineering. Accurate predictions of properties like hemolysis, solubility, and resistance to nonspecific interactions can aid in identifying peptides with desired characteristics for therapeutic or functional purposes.

\section{Data and software availability}

The necessary code (including scripts to download the datasets) used in this study can be accessed here:
\href{https://github.com/ChakradharG/PeptideBERT}{https://github.com/ChakradharG/PeptideBERT}

\section{Supporting Information}
\subsection{Optimal Hyperparameters and Training Time}

The effectiveness of our model is evident from its training time across various prediction tasks as highlighted in Table \ref{tab:traintime}. For the\textit{Nonfouling} task, the model required 58.28 minutes for training, while for\textit{Hemolysis} prediction, the training time was slightly longer at 69.28 minutes. The \textit{Solubility} prediction task demanded more extensive training, taking 116.42 minutes to converge. The hyperparameters that played a pivotal role in shaping our model's performance, are illustrated in Table \ref{tab:hyperparams}. The optimal hyperparameter values were determined after a careful fine-tuning process. An initial learning rate (Initial LR) of $1.0 * 10^{-5}$ was determined to be the optimal learning rate, managing a trade-off between rapid convergence and preventing overfitting. The model performed well with a batch size of 32, and the model configuration consisted of 12 attention heads and 12 hidden layers, each comprising 480 hidden units. To prevent overfitting, a dropout rate of 0.15 was employed between hidden layers. The learning rate scheduler, with a reduction factor of 0.1, along with the patience of 4, contributed to a more stable convergence process.

\begin{table}[t!]
    \begin{tabular}{ |c|c| } 
    \hline
    Task & Training time (minutes)  \\
    \hline
    Nonfouling & 58.28 \\
    Hemolysis & 69.28 \\
    Solubility & 116.42 \\
    \hline
    \end{tabular}
    \caption{\label{tab:traintime}Time taken to train the model on each of the 3 prediction tasks}
\end{table}

\begin{table}[t!]
    \begin{tabular}{ |c|c| } 
    \hline
    Hyperparameter & Optimal value  \\
    \hline
    Initial LR & $1.0 * 10^{-5}$ \\
    Batch Size & 32 \\
    Number of Attention Heads & 12 \\
    Number of Hidden Layers & 12 \\
    Hidden Size & 480 \\
    Hidden Layer Dropout & 0.15 \\
    LR Scheduler (factor) & 0.1 \\
    LR Scheduler (Patience) & 4 \\
    \hline
    \end{tabular}
    \caption{\label{tab:hyperparams}Hyperparameters along with their optimal values used to train PeptideBERT}
\end{table}

\subsubsection{Additional ablation studies for the Solubility and Hemolysis tasks}
In order to assess the effectiveness of different data augmentation techniques in improving the performance of our model for the Solubility task, we conducted some additional ablation studies as outlined in Table \ref{tab:augmentation2}. The ablation study involved applying \textit{Random Masking} to the training set, at different masking probabilities. The results revealed a pattern of diminishing accuracy as the augmentation level increased. Specifically, when applying a random masking probability of 0.15, our model achieved an accuracy of 68.784\%. A slight decrease in accuracy was observed when masking probability of 0.20 was applied(67.863\%). Further decrease in accuracy was observed when the masking probability was further increased to 0.30. These findings indicate the trade-off between data augmentation and model performance. Results of the ablation studies shown in Table \ref{tab:hyperparam2} shed light on how different architectural configurations influence performance outcomes. 

\begin{table}[t!]
    \begin{tabular}{ |c|c|c| } 
    \hline
    Augmentations applied & Train set size & Accuracy(\%) \\
    \hline
	random\_mask(15\%) & 29524 & 68.784 \\
	random\_mask(20\%) & 29524 & 67.863 \\
	random\_mask(30\%) & 29524 & 65.894 \\
    \hline
    \end{tabular}
    \caption{\label{tab:augmentation2}Ablation results of other augmentation techniques applied for the \textit{Solubility} task}
\end{table}

\begin{table}[t!]
    \begin{tabular}{ |c|c|c| } 
    \hline
    Hyperparameters & Value & Accuracy(\%) \\
    \hline
	num\_hidden\_layers,hidden\_dim,num\_attention\_heads  & 12,480,12  & 83.010 \\
	num\_hidden\_layers,hidden\_dim,num\_attention\_heads  & 48,560,24  & 78.865 \\
    \hline
    \end{tabular}
    \caption{\label{tab:hyperparam2}Results of the ablation studies for the \textit{Hemolysis} task}
\end{table}

\noindent The first configuration with 12 hidden layers, a hidden dimension of 480, and 12 attention heads demonstrated an accuracy of 83.010\%. On the other hand, the second configuration, characterized by a more complex architecture with 48 hidden layers, a larger hidden dimension of 560, and 24 attention heads, achieves a still commendable accuracy of 78.865\%. This indicates that while increased model depth and attention head count can potentially introduce more intricate representations in the model architecture, there exists a threshold beyond which the advantages of having more complex representations might plateau or even diminish.

\begin{acknowledgement}

This work is supported by the Center for Machine Learning in Health (CMLH) at Carnegie Mellon University and a start-up fund from Mechanical Engineering Department at CMU. 

\end{acknowledgement}

\bibliography{reference}

\providecommand{\latin}[1]{#1}
\makeatletter
\providecommand{\doi}
  {\begingroup\let\do\@makeother\dospecials
  \catcode`\{=1 \catcode`\}=2 \doi@aux}
\providecommand{\doi@aux}[1]{\endgroup\texttt{#1}}
\makeatother
\providecommand*\mcitethebibliography{\thebibliography}
\csname @ifundefined\endcsname{endmcitethebibliography}
  {\let\endmcitethebibliography\endthebibliography}{}
\begin{mcitethebibliography}{54}
\providecommand*\natexlab[1]{#1}
\providecommand*\mciteSetBstSublistMode[1]{}
\providecommand*\mciteSetBstMaxWidthForm[2]{}
\providecommand*\mciteBstWouldAddEndPuncttrue
  {\def\EndOfBibitem{\unskip.}}
\providecommand*\mciteBstWouldAddEndPunctfalse
  {\let\EndOfBibitem\relax}
\providecommand*\mciteSetBstMidEndSepPunct[3]{}
\providecommand*\mciteSetBstSublistLabelBeginEnd[3]{}
\providecommand*\EndOfBibitem{}
\mciteSetBstSublistMode{f}
\mciteSetBstMaxWidthForm{subitem}{(\alph{mcitesubitemcount})}
\mciteSetBstSublistLabelBeginEnd
  {\mcitemaxwidthsubitemform\space}
  {\relax}
  {\relax}

\bibitem[Langel \latin{et~al.}(2009)Langel, Cravatt, Graslund, Von~Heijne,
  Zorko, Land, and Niessen]{langel2009introduction}
Langel,~U.; Cravatt,~B.~F.; Graslund,~A.; Von~Heijne,~N.; Zorko,~M.; Land,~T.;
  Niessen,~S. \emph{Introduction to peptides and proteins}; CRC press,
  2009\relax
\mciteBstWouldAddEndPuncttrue
\mciteSetBstMidEndSepPunct{\mcitedefaultmidpunct}
{\mcitedefaultendpunct}{\mcitedefaultseppunct}\relax
\EndOfBibitem
\bibitem[Damodaran(2008)]{damodaran2008amino}
Damodaran,~S. Amino acids, peptides and proteins. \emph{Fennema’s food
  chemistry} \textbf{2008}, \emph{4}, 425--439\relax
\mciteBstWouldAddEndPuncttrue
\mciteSetBstMidEndSepPunct{\mcitedefaultmidpunct}
{\mcitedefaultendpunct}{\mcitedefaultseppunct}\relax
\EndOfBibitem
\bibitem[Degrado(1988)]{degrado1988design}
Degrado,~W.~F. Design of peptides and proteins. \emph{Advances in protein
  chemistry} \textbf{1988}, \emph{39}, 51--124\relax
\mciteBstWouldAddEndPuncttrue
\mciteSetBstMidEndSepPunct{\mcitedefaultmidpunct}
{\mcitedefaultendpunct}{\mcitedefaultseppunct}\relax
\EndOfBibitem
\bibitem[Voet \latin{et~al.}(2016)Voet, Voet, and Pratt]{voet2016fundamentals}
Voet,~D.; Voet,~J.~G.; Pratt,~C.~W. \emph{Fundamentals of biochemistry: life at
  the molecular level}; John Wiley \& Sons, 2016\relax
\mciteBstWouldAddEndPuncttrue
\mciteSetBstMidEndSepPunct{\mcitedefaultmidpunct}
{\mcitedefaultendpunct}{\mcitedefaultseppunct}\relax
\EndOfBibitem
\bibitem[Bodanszky(2012)]{bodanszky2012principles}
Bodanszky,~M. \emph{Principles of peptide synthesis}; Springer Science \&
  Business Media, 2012; Vol.~16\relax
\mciteBstWouldAddEndPuncttrue
\mciteSetBstMidEndSepPunct{\mcitedefaultmidpunct}
{\mcitedefaultendpunct}{\mcitedefaultseppunct}\relax
\EndOfBibitem
\bibitem[Mollaei and Farimani(2023)Mollaei, and
  Farimani]{Mollaei2023.04.16.536913}
Mollaei,~P.; Farimani,~A.~B. A Machine Learning Method to Characterize
  Conformational Changes of Amino Acids in Proteins. \emph{bioRxiv}
  \textbf{2023}, \relax
\mciteBstWouldAddEndPunctfalse
\mciteSetBstMidEndSepPunct{\mcitedefaultmidpunct}
{}{\mcitedefaultseppunct}\relax
\EndOfBibitem
\bibitem[Schulz and Schirmer(2013)Schulz, and Schirmer]{schulz2013principles}
Schulz,~G.~E.; Schirmer,~R.~H. \emph{Principles of protein structure}; Springer
  Science \& Business Media, 2013\relax
\mciteBstWouldAddEndPuncttrue
\mciteSetBstMidEndSepPunct{\mcitedefaultmidpunct}
{\mcitedefaultendpunct}{\mcitedefaultseppunct}\relax
\EndOfBibitem
\bibitem[Petsko and Ringe(2004)Petsko, and Ringe]{petsko2004protein}
Petsko,~G.~A.; Ringe,~D. \emph{Protein structure and function}; New Science
  Press, 2004\relax
\mciteBstWouldAddEndPuncttrue
\mciteSetBstMidEndSepPunct{\mcitedefaultmidpunct}
{\mcitedefaultendpunct}{\mcitedefaultseppunct}\relax
\EndOfBibitem
\bibitem[Mollaei and Barati~Farimani(2023)Mollaei, and
  Barati~Farimani]{mollaei2023activity}
Mollaei,~P.; Barati~Farimani,~A. Activity Map and Transition Pathways of G
  Protein-Coupled Receptor Revealed by Machine Learning. \emph{Journal of
  Chemical Information and Modeling} \textbf{2023}, \emph{63}, 2296--2304\relax
\mciteBstWouldAddEndPuncttrue
\mciteSetBstMidEndSepPunct{\mcitedefaultmidpunct}
{\mcitedefaultendpunct}{\mcitedefaultseppunct}\relax
\EndOfBibitem
\bibitem[Yadav \latin{et~al.}(2022)Yadav, Mollaei, Cao, Wang, and
  Farimani]{yadav2022prediction}
Yadav,~P.; Mollaei,~P.; Cao,~Z.; Wang,~Y.; Farimani,~A.~B. Prediction of GPCR
  activity using machine learning. \emph{Computational and Structural
  Biotechnology Journal} \textbf{2022}, \emph{20}, 2564--2573\relax
\mciteBstWouldAddEndPuncttrue
\mciteSetBstMidEndSepPunct{\mcitedefaultmidpunct}
{\mcitedefaultendpunct}{\mcitedefaultseppunct}\relax
\EndOfBibitem
\bibitem[Varanko \latin{et~al.}(2020)Varanko, Saha, and
  Chilkoti]{varanko2020recent}
Varanko,~A.; Saha,~S.; Chilkoti,~A. Recent trends in protein and peptide-based
  biomaterials for advanced drug delivery. \emph{Advanced drug delivery
  reviews} \textbf{2020}, \emph{156}, 133--187\relax
\mciteBstWouldAddEndPuncttrue
\mciteSetBstMidEndSepPunct{\mcitedefaultmidpunct}
{\mcitedefaultendpunct}{\mcitedefaultseppunct}\relax
\EndOfBibitem
\bibitem[Dunn(2015)]{dunn2015peptide}
Dunn,~B.~M. \emph{Peptide chemistry and drug design}; Wiley Online Library,
  2015\relax
\mciteBstWouldAddEndPuncttrue
\mciteSetBstMidEndSepPunct{\mcitedefaultmidpunct}
{\mcitedefaultendpunct}{\mcitedefaultseppunct}\relax
\EndOfBibitem
\bibitem[Schueler-Furman \latin{et~al.}(2017)Schueler-Furman, London, and
  Schueler-Furman]{schueler2017modeling}
Schueler-Furman,~O.; London,~N.; Schueler-Furman, \emph{Modeling
  peptide-protein interactions}; Springer, 2017\relax
\mciteBstWouldAddEndPuncttrue
\mciteSetBstMidEndSepPunct{\mcitedefaultmidpunct}
{\mcitedefaultendpunct}{\mcitedefaultseppunct}\relax
\EndOfBibitem
\bibitem[Ponder(1948)]{ponder1948hemolysis}
Ponder,~E. \emph{Hemolysis and related phenomena}; Saunders, 1948\relax
\mciteBstWouldAddEndPuncttrue
\mciteSetBstMidEndSepPunct{\mcitedefaultmidpunct}
{\mcitedefaultendpunct}{\mcitedefaultseppunct}\relax
\EndOfBibitem
\bibitem[Harding and Reynolds(2014)Harding, and Reynolds]{harding2014combating}
Harding,~J.~L.; Reynolds,~M.~M. Combating medical device fouling. \emph{Trends
  in biotechnology} \textbf{2014}, \emph{32}, 140--146\relax
\mciteBstWouldAddEndPuncttrue
\mciteSetBstMidEndSepPunct{\mcitedefaultmidpunct}
{\mcitedefaultendpunct}{\mcitedefaultseppunct}\relax
\EndOfBibitem
\bibitem[Yu \latin{et~al.}(2011)Yu, Zhang, Wang, Brash, and Chen]{yu2011anti}
Yu,~Q.; Zhang,~Y.; Wang,~H.; Brash,~J.; Chen,~H. Anti-fouling bioactive
  surfaces. \emph{Acta biomaterialia} \textbf{2011}, \emph{7}, 1550--1557\relax
\mciteBstWouldAddEndPuncttrue
\mciteSetBstMidEndSepPunct{\mcitedefaultmidpunct}
{\mcitedefaultendpunct}{\mcitedefaultseppunct}\relax
\EndOfBibitem
\bibitem[Sarma \latin{et~al.}(2018)Sarma, Wong, Lynch, and
  Pettitt]{sarma2018peptide}
Sarma,~R.; Wong,~K.-Y.; Lynch,~G.~C.; Pettitt,~B.~M. Peptide solubility limits:
  backbone and side-chain interactions. \emph{The Journal of Physical Chemistry
  B} \textbf{2018}, \emph{122}, 3528--3539\relax
\mciteBstWouldAddEndPuncttrue
\mciteSetBstMidEndSepPunct{\mcitedefaultmidpunct}
{\mcitedefaultendpunct}{\mcitedefaultseppunct}\relax
\EndOfBibitem
\bibitem[Fosgerau and Hoffmann(2015)Fosgerau, and
  Hoffmann]{fosgerau2015peptide}
Fosgerau,~K.; Hoffmann,~T. Peptide therapeutics: current status and future
  directions. \emph{Drug discovery today} \textbf{2015}, \emph{20},
  122--128\relax
\mciteBstWouldAddEndPuncttrue
\mciteSetBstMidEndSepPunct{\mcitedefaultmidpunct}
{\mcitedefaultendpunct}{\mcitedefaultseppunct}\relax
\EndOfBibitem
\bibitem[Cherkasov \latin{et~al.}(2014)Cherkasov, Muratov, Fourches, Varnek,
  Baskin, Cronin, Dearden, Gramatica, Martin, Todeschini, Consonni, Kuz’min,
  Cramer, Benigni, Yang, Rathman, Terfloth, Gasteiger, Richard, and
  Tropsha]{QSAR2014modelling}
Cherkasov,~A. \latin{et~al.}  QSAR Modeling: Where Have You Been? Where Are You
  Going To? \emph{Journal of Medicinal Chemistry} \textbf{2014}, \emph{57},
  4977--5010, PMID: 24351051\relax
\mciteBstWouldAddEndPuncttrue
\mciteSetBstMidEndSepPunct{\mcitedefaultmidpunct}
{\mcitedefaultendpunct}{\mcitedefaultseppunct}\relax
\EndOfBibitem
\bibitem[Deng \latin{et~al.}(2017)Deng, Ni, Zhai, Tang, Tan, Yan, Deng, and
  Yin]{newquantitativestructure}
Deng,~B.; Ni,~X.; Zhai,~Z.; Tang,~T.; Tan,~C.; Yan,~Y.; Deng,~J.; Yin,~Y. New
  Quantitative Structure–Activity Relationship Model for
  Angiotensin-Converting Enzyme Inhibitory Dipeptides Based on Integrated
  Descriptors. \emph{Journal of Agricultural and Food Chemistry} \textbf{2017},
  \emph{65}, 9774--9781, PMID: 28984136\relax
\mciteBstWouldAddEndPuncttrue
\mciteSetBstMidEndSepPunct{\mcitedefaultmidpunct}
{\mcitedefaultendpunct}{\mcitedefaultseppunct}\relax
\EndOfBibitem
\bibitem[Wang \latin{et~al.}(2020)Wang, Russo, Liu, Zhou, Zhu, and
  Zhang]{ACE-I}
Wang,~Y.-T.; Russo,~D.~P.; Liu,~C.; Zhou,~Q.; Zhu,~H.; Zhang,~Y.-H. Predictive
  Modeling of Angiotensin I-Converting Enzyme Inhibitory Peptides Using Various
  Machine Learning Approaches. \emph{Journal of Agricultural and Food
  Chemistry} \textbf{2020}, \emph{68}, 12132--12140, PMID: 32915574\relax
\mciteBstWouldAddEndPuncttrue
\mciteSetBstMidEndSepPunct{\mcitedefaultmidpunct}
{\mcitedefaultendpunct}{\mcitedefaultseppunct}\relax
\EndOfBibitem
\bibitem[Guan~X.(2019)]{ACE-1study}
Guan~X.,~L.~J. QSAR Study of Angiotensin I-Converting Enzyme Inhibitory
  Peptides Using SVHEHS Descriptor and OSC-SVM. \textbf{2019}, \relax
\mciteBstWouldAddEndPunctfalse
\mciteSetBstMidEndSepPunct{\mcitedefaultmidpunct}
{}{\mcitedefaultseppunct}\relax
\EndOfBibitem
\bibitem[Vishnepolsky \latin{et~al.}(2018)Vishnepolsky, Gabrielian, Rosenthal,
  Hurt, Tartakovsky, Managadze, Grigolava, Makhatadze, and
  Pirtskhalava]{Gramnegative}
Vishnepolsky,~B.; Gabrielian,~A.; Rosenthal,~A.; Hurt,~D.~E.; Tartakovsky,~M.;
  Managadze,~G.; Grigolava,~M.; Makhatadze,~G.~I.; Pirtskhalava,~M. Predictive
  Model of Linear Antimicrobial Peptides Active against Gram-Negative Bacteria.
  \emph{Journal of Chemical Information and Modeling} \textbf{2018}, \emph{58},
  1141--1151, PMID: 29716188\relax
\mciteBstWouldAddEndPuncttrue
\mciteSetBstMidEndSepPunct{\mcitedefaultmidpunct}
{\mcitedefaultendpunct}{\mcitedefaultseppunct}\relax
\EndOfBibitem
\bibitem[Barrett \latin{et~al.}(2018)Barrett, Jiang, and White]{Gramnegative2}
Barrett,~R.; Jiang,~S.; White,~A. J.~P. Classifying antimicrobial and
  multifunctional peptides with bayesian network models. \emph{Peptide Science}
  \textbf{2018}, \emph{110}\relax
\mciteBstWouldAddEndPuncttrue
\mciteSetBstMidEndSepPunct{\mcitedefaultmidpunct}
{\mcitedefaultendpunct}{\mcitedefaultseppunct}\relax
\EndOfBibitem
\bibitem[Das \latin{et~al.}(2021)Das, Sercu, Wadhawan, Padhi, Gehrmann,
  Cipcigan, Chenthamarakshan, Strobelt, Dos~Santos, Chen, Yang, Tan, Hedrick,
  Crain, and Mojsilovic]{Gramnegative3}
Das,~P.; Sercu,~T.; Wadhawan,~K.; Padhi,~I.; Gehrmann,~S.; Cipcigan,~F.;
  Chenthamarakshan,~V.; Strobelt,~H.; Dos~Santos,~C.; Chen,~P.-Y.; Yang,~Y.~Y.;
  Tan,~J. P.~K.; Hedrick,~J.; Crain,~J.; Mojsilovic,~A. Accelerated
  antimicrobial discovery via deep generative models and molecular dynamics
  simulations. \emph{Nature biomedical engineering} \textbf{2021}, \emph{5},
  613—623\relax
\mciteBstWouldAddEndPuncttrue
\mciteSetBstMidEndSepPunct{\mcitedefaultmidpunct}
{\mcitedefaultendpunct}{\mcitedefaultseppunct}\relax
\EndOfBibitem
\bibitem[Chen \latin{et~al.}(2018)Chen, Chen, Yao, and Li]{antioxidants}
Chen,~N.; Chen,~J.; Yao,~B.; Li,~Z. QSAR Study on Antioxidant Tripeptides and
  the Antioxidant Activity of the Designed Tripeptides in Free Radical Systems.
  \emph{Molecules} \textbf{2018}, \emph{23}\relax
\mciteBstWouldAddEndPuncttrue
\mciteSetBstMidEndSepPunct{\mcitedefaultmidpunct}
{\mcitedefaultendpunct}{\mcitedefaultseppunct}\relax
\EndOfBibitem
\bibitem[Deng \latin{et~al.}(2019)Deng, Long, Tang, Ni, Chen, Yang, Zhang, Cao,
  Cao, Zeng, and Yi]{antioxidants1}
Deng,~B.; Long,~H.; Tang,~T.; Ni,~X.; Chen,~J.; Yang,~G.; Zhang,~F.; Cao,~R.;
  Cao,~D.; Zeng,~M.; Yi,~L. Quantitative Structure-Activity Relationship Study
  of Antioxidant Tripeptides Based on Model Population Analysis.
  \emph{International journal of molecular sciences} \textbf{2019}, \emph{20},
  E995\relax
\mciteBstWouldAddEndPuncttrue
\mciteSetBstMidEndSepPunct{\mcitedefaultmidpunct}
{\mcitedefaultendpunct}{\mcitedefaultseppunct}\relax
\EndOfBibitem
\bibitem[Olsen \latin{et~al.}(2020)Olsen, Yesiltas, Marin, Pertseva,
  García-Moreno, Gregersen, Overgaard, Jacobsen, Lund, Hansen, and
  Marcatili]{antioxidants2}
Olsen,~T.~H.; Yesiltas,~B.; Marin,~F.~I.; Pertseva,~M.; García-Moreno,~P.~J.;
  Gregersen,~S.; Overgaard,~M.~T.; Jacobsen,~C.; Lund,~O.; Hansen,~E.~B.;
  Marcatili,~P. AnOxPePred: using deep learning for the prediction of
  antioxidative properties of peptides. \emph{Scientific reports}
  \textbf{2020}, \emph{10}, 21471\relax
\mciteBstWouldAddEndPuncttrue
\mciteSetBstMidEndSepPunct{\mcitedefaultmidpunct}
{\mcitedefaultendpunct}{\mcitedefaultseppunct}\relax
\EndOfBibitem
\bibitem[Madani \latin{et~al.}(2021)Madani, Lin, and Tarakanova]{DSResol}
Madani,~M.; Lin,~K.; Tarakanova,~A. DSResSol: A sequence-based solubility
  predictor created with Dilated Squeeze Excitation Residual Networks.
  \emph{bioRxiv} \textbf{2021}, \relax
\mciteBstWouldAddEndPunctfalse
\mciteSetBstMidEndSepPunct{\mcitedefaultmidpunct}
{}{\mcitedefaultseppunct}\relax
\EndOfBibitem
\bibitem[Khurana \latin{et~al.}(2018)Khurana, Rawi, Kunji, Chuang, Bensmail,
  and Mall]{DeepSol}
Khurana,~S.; Rawi,~R.; Kunji,~K.; Chuang,~G.-Y.; Bensmail,~H.; Mall,~R.
  {DeepSol: a deep learning framework for sequence-based protein solubility
  prediction}. \emph{Bioinformatics} \textbf{2018}, \emph{34}, 2605--2613\relax
\mciteBstWouldAddEndPuncttrue
\mciteSetBstMidEndSepPunct{\mcitedefaultmidpunct}
{\mcitedefaultendpunct}{\mcitedefaultseppunct}\relax
\EndOfBibitem
\bibitem[Hon \latin{et~al.}(2021)Hon, Marusiak, Martinek, Kunka, Zendulka,
  Bednar, and Damborsky]{Soluprot}
Hon,~J.; Marusiak,~M.; Martinek,~T.; Kunka,~A.; Zendulka,~J.; Bednar,~D.;
  Damborsky,~J. {SoluProt: prediction of soluble protein expression in
  Escherichia coli}. \emph{Bioinformatics} \textbf{2021}, \emph{37},
  23--28\relax
\mciteBstWouldAddEndPuncttrue
\mciteSetBstMidEndSepPunct{\mcitedefaultmidpunct}
{\mcitedefaultendpunct}{\mcitedefaultseppunct}\relax
\EndOfBibitem
\bibitem[Hebditch \latin{et~al.}(2017)Hebditch, Carballo-Amador, Charonis,
  Curtis, and Warwicker]{proteinsol}
Hebditch,~M.; Carballo-Amador,~M.~A.; Charonis,~S.; Curtis,~R.; Warwicker,~J.
  {Protein–Sol: a web tool for predicting protein solubility from sequence}.
  \emph{Bioinformatics} \textbf{2017}, \emph{33}, 3098--3100\relax
\mciteBstWouldAddEndPuncttrue
\mciteSetBstMidEndSepPunct{\mcitedefaultmidpunct}
{\mcitedefaultendpunct}{\mcitedefaultseppunct}\relax
\EndOfBibitem
\bibitem[Ansari and White(2023)Ansari, and White]{MahLooL}
Ansari,~M.; White,~A.~D. Serverless prediction of peptide properties with
  recurrent neural networks. \emph{Journal of Chemical Information and
  Modeling} \textbf{2023}, \emph{63}, 2546--2553\relax
\mciteBstWouldAddEndPuncttrue
\mciteSetBstMidEndSepPunct{\mcitedefaultmidpunct}
{\mcitedefaultendpunct}{\mcitedefaultseppunct}\relax
\EndOfBibitem
\bibitem[Timmons and Hewage(2020)Timmons, and Hewage]{HAPPENN}
Timmons,~P.~B.; Hewage,~C.~M. HAPPENN is a novel tool for hemolytic activity
  prediction for therapeutic peptides which employs neural networks.
  \emph{Scientific reports} \textbf{2020}, \emph{10}, 10869\relax
\mciteBstWouldAddEndPuncttrue
\mciteSetBstMidEndSepPunct{\mcitedefaultmidpunct}
{\mcitedefaultendpunct}{\mcitedefaultseppunct}\relax
\EndOfBibitem
\bibitem[Vaswani \latin{et~al.}(2017)Vaswani, Shazeer, Parmar, Uszkoreit,
  Jones, Gomez, Kaiser, and Polosukhin]{vaswani2017attention}
Vaswani,~A.; Shazeer,~N.; Parmar,~N.; Uszkoreit,~J.; Jones,~L.; Gomez,~A.~N.;
  Kaiser,~{\L}.; Polosukhin,~I. Attention is all you need. \emph{Advances in
  neural information processing systems} \textbf{2017}, \emph{30}\relax
\mciteBstWouldAddEndPuncttrue
\mciteSetBstMidEndSepPunct{\mcitedefaultmidpunct}
{\mcitedefaultendpunct}{\mcitedefaultseppunct}\relax
\EndOfBibitem
\bibitem[Devlin \latin{et~al.}(2018)Devlin, Chang, Lee, and
  Toutanova]{devlin2018bert}
Devlin,~J.; Chang,~M.-W.; Lee,~K.; Toutanova,~K. Bert: Pre-training of deep
  bidirectional transformers for language understanding. \emph{arXiv preprint
  arXiv:1810.04805} \textbf{2018}, \relax
\mciteBstWouldAddEndPunctfalse
\mciteSetBstMidEndSepPunct{\mcitedefaultmidpunct}
{}{\mcitedefaultseppunct}\relax
\EndOfBibitem
\bibitem[Brown \latin{et~al.}(2020)Brown, Mann, Ryder, Subbiah, Kaplan,
  Dhariwal, Neelakantan, Shyam, Sastry, Askell, \latin{et~al.}
  others]{brown2020language}
Brown,~T.; Mann,~B.; Ryder,~N.; Subbiah,~M.; Kaplan,~J.~D.; Dhariwal,~P.;
  Neelakantan,~A.; Shyam,~P.; Sastry,~G.; Askell,~A., \latin{et~al.}  Language
  models are few-shot learners. \emph{Advances in neural information processing
  systems} \textbf{2020}, \emph{33}, 1877--1901\relax
\mciteBstWouldAddEndPuncttrue
\mciteSetBstMidEndSepPunct{\mcitedefaultmidpunct}
{\mcitedefaultendpunct}{\mcitedefaultseppunct}\relax
\EndOfBibitem
\bibitem[Elnaggar \latin{et~al.}(2020)Elnaggar, Heinzinger, Dallago, Rehawi,
  Wang, Jones, Gibbs, Feher, Angerer, Steinegger, BHOWMIK, and Rost]{protbert}
Elnaggar,~A.; Heinzinger,~M.; Dallago,~C.; Rehawi,~G.; Wang,~Y.; Jones,~L.;
  Gibbs,~T.; Feher,~T.; Angerer,~C.; Steinegger,~M.; BHOWMIK,~D.; Rost,~B.
  ProtTrans: Towards Cracking the Language of Life{\textquoteright}s Code
  Through Self-Supervised Deep Learning and High Performance Computing.
  \emph{bioRxiv} \textbf{2020}, \relax
\mciteBstWouldAddEndPunctfalse
\mciteSetBstMidEndSepPunct{\mcitedefaultmidpunct}
{}{\mcitedefaultseppunct}\relax
\EndOfBibitem
\bibitem[Marqus~S(2017)]{AMPs}
Marqus~S,~P.~T.,~Pirogova~E Evaluation of the use of therapeutic peptides for
  cancer treatment. \textbf{2017}, \relax
\mciteBstWouldAddEndPunctfalse
\mciteSetBstMidEndSepPunct{\mcitedefaultmidpunct}
{}{\mcitedefaultseppunct}\relax
\EndOfBibitem
\bibitem[Kunda(2020)]{AMPs1}
Kunda,~N.~K. Antimicrobial peptides as novel therapeutics for non-small cell
  lung cancer. \emph{Drug Discovery Today} \textbf{2020}, \emph{25},
  238--247\relax
\mciteBstWouldAddEndPuncttrue
\mciteSetBstMidEndSepPunct{\mcitedefaultmidpunct}
{\mcitedefaultendpunct}{\mcitedefaultseppunct}\relax
\EndOfBibitem
\bibitem[Gogoladze \latin{et~al.}(2014)Gogoladze, Grigolava, Vishnepolsky,
  Chubinidze, Duroux, Lefranc, and Pirtskhalava]{DBAASP}
Gogoladze,~G.; Grigolava,~M.; Vishnepolsky,~B.; Chubinidze,~M.; Duroux,~P.;
  Lefranc,~M.-P.; Pirtskhalava,~M. DBAASP: database of antimicrobial activity
  and structure of peptides. \emph{FEMS microbiology letters} \textbf{2014},
  \emph{357}, 63—68\relax
\mciteBstWouldAddEndPuncttrue
\mciteSetBstMidEndSepPunct{\mcitedefaultmidpunct}
{\mcitedefaultendpunct}{\mcitedefaultseppunct}\relax
\EndOfBibitem
\bibitem[Smialowski \latin{et~al.}(2012)Smialowski, Doose, Torkler, Kaufmann,
  and Frishman]{PROSO}
Smialowski,~P.; Doose,~G.; Torkler,~P.; Kaufmann,~S.; Frishman,~D. PROSO II--a
  new method for protein solubility prediction. \emph{The FEBS journal}
  \textbf{2012}, \emph{279}, 2192—2200\relax
\mciteBstWouldAddEndPuncttrue
\mciteSetBstMidEndSepPunct{\mcitedefaultmidpunct}
{\mcitedefaultendpunct}{\mcitedefaultseppunct}\relax
\EndOfBibitem
\bibitem[Berman \latin{et~al.}(2008)Berman, Westbrook, Gabanyi, Tao, Shah,
  Kouranov, Schwede, Arnold, Kiefer, Bordoli, Kopp, Podvinec, Adams, Carter,
  Minor, Nair, and Baer]{PRSISGKB}
Berman,~H.~M. \latin{et~al.}  {The protein structure initiative structural
  genomics knowledgebase}. \emph{Nucleic Acids Research} \textbf{2008},
  \emph{37}, D365--D368\relax
\mciteBstWouldAddEndPuncttrue
\mciteSetBstMidEndSepPunct{\mcitedefaultmidpunct}
{\mcitedefaultendpunct}{\mcitedefaultseppunct}\relax
\EndOfBibitem
\bibitem[{Barrett} \latin{et~al.}(2018){Barrett}, {Jiang}, and
  {White}]{nonfouling}
{Barrett},~R.; {Jiang},~S.; {White},~A.~D. {Classifying Antimicrobial and
  Multifunctional Peptides with Bayesian Network Models}. \emph{arXiv e-prints}
  \textbf{2018}, arXiv:1804.06327\relax
\mciteBstWouldAddEndPuncttrue
\mciteSetBstMidEndSepPunct{\mcitedefaultmidpunct}
{\mcitedefaultendpunct}{\mcitedefaultseppunct}\relax
\EndOfBibitem
\bibitem[White \latin{et~al.}(2012)White, Nowinski, Huang, Keefe, Sun, and
  Jiang]{C2SC21135A}
White,~A.~D.; Nowinski,~A.~K.; Huang,~W.; Keefe,~A.~J.; Sun,~F.; Jiang,~S.
  Decoding nonspecific interactions from nature. \emph{Chem. Sci.}
  \textbf{2012}, \emph{3}, 3488--3494\relax
\mciteBstWouldAddEndPuncttrue
\mciteSetBstMidEndSepPunct{\mcitedefaultmidpunct}
{\mcitedefaultendpunct}{\mcitedefaultseppunct}\relax
\EndOfBibitem
\bibitem[White \latin{et~al.}(2012)White, Huang, and Jiang]{WHITE20122484}
White,~A.; Huang,~W.; Jiang,~S. Role of Nonspecific Interactions in Molecular
  Chaperones through Model-Based Bioinformatics. \emph{Biophysical Journal}
  \textbf{2012}, \emph{103}, 2484--2491\relax
\mciteBstWouldAddEndPuncttrue
\mciteSetBstMidEndSepPunct{\mcitedefaultmidpunct}
{\mcitedefaultendpunct}{\mcitedefaultseppunct}\relax
\EndOfBibitem
\bibitem[Devlin \latin{et~al.}(2019)Devlin, Chang, Lee, and Toutanova]{bert}
Devlin,~J.; Chang,~M.-W.; Lee,~K.; Toutanova,~K. BERT: Pre-training of Deep
  Bidirectional Transformers for Language Understanding. 2019\relax
\mciteBstWouldAddEndPuncttrue
\mciteSetBstMidEndSepPunct{\mcitedefaultmidpunct}
{\mcitedefaultendpunct}{\mcitedefaultseppunct}\relax
\EndOfBibitem
\bibitem[Suzek \latin{et~al.}(2015)Suzek, Wang, Huang, McGarvey, Wu, and
  Consortium]{uniref100}
Suzek,~B.~E.; Wang,~Y.; Huang,~H.; McGarvey,~P.~B.; Wu,~C.~H.; Consortium,~U.
  UniRef clusters: a comprehensive and scalable alternative for improving
  sequence similarity searches. \emph{Bioinformatics} \textbf{2015}, \emph{31},
  926--932\relax
\mciteBstWouldAddEndPuncttrue
\mciteSetBstMidEndSepPunct{\mcitedefaultmidpunct}
{\mcitedefaultendpunct}{\mcitedefaultseppunct}\relax
\EndOfBibitem
\bibitem[Hasan \latin{et~al.}(2020)Hasan, Schaduangrat, Basith, Lee,
  Shoombuatong, and Manavalan]{HLPpred}
Hasan,~M.~M.; Schaduangrat,~N.; Basith,~S.; Lee,~G.; Shoombuatong,~W.;
  Manavalan,~B. {HLPpred-Fuse: improved and robust prediction of hemolytic
  peptide and its activity by fusing multiple feature representation}.
  \emph{Bioinformatics} \textbf{2020}, \emph{36}, 3350--3356\relax
\mciteBstWouldAddEndPuncttrue
\mciteSetBstMidEndSepPunct{\mcitedefaultmidpunct}
{\mcitedefaultendpunct}{\mcitedefaultseppunct}\relax
\EndOfBibitem
\bibitem[Capecchi \latin{et~al.}(2021)Capecchi, Cai, Personne, Köhler, van
  Delden, and Reymond]{onehots}
Capecchi,~A.; Cai,~X.; Personne,~H.; Köhler,~T.; van Delden,~C.;
  Reymond,~J.-L. Machine learning designs non-hemolytic antimicrobial peptides.
  \emph{Chem. Sci.} \textbf{2021}, \emph{12}, 9221--9232\relax
\mciteBstWouldAddEndPuncttrue
\mciteSetBstMidEndSepPunct{\mcitedefaultmidpunct}
{\mcitedefaultendpunct}{\mcitedefaultseppunct}\relax
\EndOfBibitem
\bibitem[Schwaller \latin{et~al.}(2019)Schwaller, Laino, Gaudin, Bolgar, Bekas,
  and Lee]{schwaller1}
Schwaller,~P.; Laino,~T.; Gaudin,~T.; Bolgar,~P.; Bekas,~C.; Lee,~A.~A.
  Molecular transformer – a model for uncertainty-calibrated chemical
  reaction prediction. \emph{ACS central science} \textbf{2019}, \emph{5}\relax
\mciteBstWouldAddEndPuncttrue
\mciteSetBstMidEndSepPunct{\mcitedefaultmidpunct}
{\mcitedefaultendpunct}{\mcitedefaultseppunct}\relax
\EndOfBibitem
\bibitem[Schwaller \latin{et~al.}(2020)Schwaller, Probst, Vaucher, Nair,
  Kreutter, Laino, and Reymond]{schwaller2}
Schwaller,~P.; Probst,~D.; Vaucher,~A.~C.; Nair,~V.~H.; Kreutter,~D.;
  Laino,~T.; Reymond,~J.-L. Mapping the space of chemical reactions using
  attention-based neural networks. \emph{Nature Machine Intelligence}
  \textbf{2020}, \emph{3}\relax
\mciteBstWouldAddEndPuncttrue
\mciteSetBstMidEndSepPunct{\mcitedefaultmidpunct}
{\mcitedefaultendpunct}{\mcitedefaultseppunct}\relax
\EndOfBibitem
\bibitem[Maaten and Hinton(2008)Maaten, and Hinton]{Maaten1990tsne}
Maaten,~L. v.~d.; Hinton,~G. Visualizing Data using t-SNE. \emph{Journal of
  Machine Learning Research} \textbf{2008}, 2579--2605\relax
\mciteBstWouldAddEndPuncttrue
\mciteSetBstMidEndSepPunct{\mcitedefaultmidpunct}
{\mcitedefaultendpunct}{\mcitedefaultseppunct}\relax
\EndOfBibitem
\end{mcitethebibliography}

\end{document}